\documentstyle[preprint,aps,prb]{revtex}  
\begin{document}
\tighten
\title{On two channel flavor anisotropic and one channel compactified Kondo models}
\author{Jinwu  Ye}
\address{
Physics Laboratories, Harvard  University, Cambridge, MA, 02138   \\
and Department of Physics and Astronomy, Johns Hopkins University,
Baltimore, MD, 21218 }
\date{\today}
\maketitle
\begin{abstract}
 We reinvestigate the two channel flavor anisotropic model (2CFAK)
 and one channel compacitified Kondo model (1CCK).
 For these two models, all {\em the possible} fixed points and their symmetries
 are identified; the finite size spectra, the electron conductivity and pairing
 susceptibility are calculated.
 It is shown that the only non-fermi liquid (NFL) fixed point of the 2CFAK is
 the NFL of the two channel Kondo model (2CK) with the symmetry $ O(3) \times O(5) $.
 Any flavor anisotropies between the two channels drive the system to the
 fermi-liquid (FL) fixed point
 with the symmetry $ O(4) \times O(4) $ where one of the two channels suffers the
 phase shift $ \pi/2 $ and the other remains free.
 The NFL fixed point of the 1CCK has the symmetry $ O(3) \times O(1) $ and has the same
 thermodynamics as the NFL fixed point of the 2CK. However, {\em in contrast to} the 2CK,
 its conductivity shows $ T^{2} $ behavior and
 there is {\em no} pairing susceptibility divergence.
 Any anisotropies between the spin and isospin sectors drive the system to the
 FL fixed point with the symmetry $ O(4) $ where the electrons suffer the
 phase shift $ \pi/2 $.
 The connection and differences between the two models are explicitly demonstrated.
 The recent conjectures and claims on the NFL behaviors of the two models are commented.
\end{abstract}
\narrowtext

\section{Introduction}
 Extensive attention has been lavished on the overscreened multichannel
 Kondo model after the discover of its non-fermi liquid (NFL) behavior by
 Nozi\'{e}res and Blandin (NB) \cite{blandin}.
 NB also pointed out that lattice effects 
 in real metals will cause the anisotropy between the two flavor channels
 and that in the low temperature limit,
 the impurity is totally screened by the strong coupling channel
 with the weak coupling channel unaffected. Using Numerical
 Renormalization Group (NRG), Ref.\cite{cox2} confirmed NB's conjecture.
 Using Conformal Field Theory (CFT), Ref.\cite{affleck}
  found a relevant dimension 1/2 operator in the flavor sector near
  the 2 channel Kondo (2CK) fixed point and
  suggested the system flows to the Fermi-liquid (FL) fixed point
  pointed out by NB. Using Yuval-Anderson's approach, Ref.\cite{gogolin}
 found a solvable line and calculated the exact crossover free energy
  function from the 2CK fixed point to the FL fixed point
  along this solvable line.

  It is known that in the large $ U $ limit, the ordinary one channel symmetric
  Anderson impurity model(AIM) can be mapped to the one channel Kondo model. However, as shown
  by Ref.\cite{com,coleman}, if the original $ O(4) $ symmetry of the AIM
  is broken to $ O(3) \times O(1) $, in the strong coupling limit, the AIM is mapped to
  the one channel compactified Kondo model (1CCK) where the impurity spin couples
  to both the spin and the isospin(charge) currents of the one channel conduction electrons.

  Recently, Andrei and Jerez \cite{andrei}, using Bethe Ansatz,
  reinvestigated the 2CFAK and conjectured that the 2CFAK flow to some
  new NFL fixed points.  Coleman and Schofield  \cite{coleman}, using strong
  coupling method, reinvestigated the 1CCK and
  claimed the system flows to another kind of non-Fermi liquid fixed point
  which, similar to 1-dim Luttinger liquid,
  has the same thermodynamics as fermi liquid but different excitation
  spectrum. Moreover, they claimed that the 1CCK has exactly the same low energy excitations as 
  those of the 2CFAK, therefore concluded that their results also apply
  to the 2CFAK.

 So far, Bethe Ansatz can only calculate thermodynamic quantities
 of multichannel Kondo models, the correlation functions are needed to resolve if the fixed 
 points are NFL or FL.
 It is important to point out that the charge degrees of freedom of the original 
 model being removed, the 1CCK in Ref.\cite{com,coleman}
 has completely different transport properties, correlation functions and excitation
 spectrums than the original 2CFAK, although it do
 share the same thermodynamic properties as the 2CFAK.

 As emphasized by AL \cite{review}, although the boundary interactions
 only happen in the spin sector;
 the spin, flavor and charge degree of freedoms
 are {\em not} totally decoupled, there is a constraint( or gluing condition)
 to describe precisely how these degree of freedoms are combined at different
 boundary fixed points, the finite size spectrum is determined by this
 gluing condition. The boundary operator contents and the scaling
 dimensions of all the boundary operators are also given by the
 gluing condition. However, in order to find the gluing conditions at the
 intermediate coupling fixed points, the fusion rules should be identified which are usually
 difficult in Non-Abelian bosonization approach.
 For 4 pieces of bulk fermions, the non-interacting theory possesses $ SO(8) $ symmetry,
 Maldacena and Ludwig (MS) \cite{ludwig} showed that finding the gluing conditions at
 the fixed points are exactly equivalent
 to finding the boundary conditions of the fermions at the fixed points;
 the CFT describing the fixed points are simply free chiral bosons with the boundary conditions.
 In Ref.\cite{powerful}, the author developed a simple and powerful method to study
 certain class of quantum impurity models. The method can quickly identify all the
 possible boundary fixed points and their {\em maximum } symmetry, therefore avoid
 the difficulty of finding the fusion rules, it can also
 demonstrate the physical picture at the boundary explicitly.  In this paper, we apply
 the method to study the two models.  All {\em the possible} fixed points and their symmetries
 are identified; the finite size spectra, the electron conductivity and pairing
 susceptibility are calculated.  All the leading and subleading
 irrelevant operators are identified, their corrections to the correlation functions
 are evaluated.  In section II, {\em Taking all the degrees of freedom into
 account}, We show  that the only NFL fixed point of the 2CFAK is
 the NFL fixed point of the 2CK with the symmetry $ O(3) \times O(5) $.
 Any flavor anisotropies between the two channels drive the system to the
 fermi-liquid (FL) fixed point
 with the symmetry $ O(4) \times O(4) $ where one of the two channels suffers the
 phase shift $ \pi/2 $ and the other remains free. 
 The conventional wisdom about the 2CFAK is rigorously shown to be correct.
 In section III, we repeat the same program to the 1CCK. We find that
 the NFL fixed point of the 1CCK has the symmetry $ O(3) \times O(1) $ and has the same
 thermodynamics as the NFL fixed point of the 2CK. The finite size spectrum
 is listed and compared with that of the 2CK. However, {\em in contrast to} the 2CK,
 its conductivity shows $ T^{2} $ bahaviour and
 there is {\em no} pairing susceptibility enhancement.
 Any anisotropies between the spin and isospin sectors drive the system to the
 FL fixed point with the symmetry $ O(4) $ where the electrons suffer the
 phase shift $ \pi/2 $. The finite size spectrum of this FL fixed point is also listed and
 compared with that of the 2CFAK. In section IV, we conclude and propose some
 open questions. Finally,
 in the appendix, we study the stable FL fixed point of the 2CFAK using Non-Abelian bosonization
 and compare with the Abelian bosonization calculations done in section II.

\section{The two channel flavor anisotropic Kondo model}
 The Hamiltonian of the 2CFAK is:
 \begin{eqnarray}
 H &= & i v_{F} \int^{\infty}_{-\infty} dx 
   \psi^{\dagger}_{i \alpha }(x) \frac{d \psi_{i \alpha }(x)}{dx}
   + \sum_{a=x,y,z} \lambda^{a} ( J^{a}_{1}(0)+J^{a}_{2}(0) )  S^{a} 
   + \sum_{a=x,y,z} \alpha^{a}  
    (J^{a}_{1}(0)-J^{a}_{2}(0)) S^{a}       \nonumber \\
  & + & h ( \int dx J^{z}_{s}(x) + S^{z} )
\label{kondob}
 \end{eqnarray}
   where $ J^{a}_{i}(x) =\frac{1}{2} \psi^{\dagger}_{i \alpha }(x)
   \sigma^{a}_{\alpha \beta} \psi_{i \beta }(x) $ 
   are the spin  currents of the channel $ i=1,2 $  conduction electrons respectively.
     $ \alpha^{a}
   =0, \pm \lambda^{a} $ correspond to the 2CK and the one channel
   Kondo model respectively.
   If $ \lambda^{a} =\lambda, \alpha^{a} =\alpha \neq 0 $ , the above Hamiltonian
   breaks  $ SU(2)_{s}  \times SU(2)_{f}  \times U(1)_{c} $ symmetry of
    the 2CK to $ SU(2)_{s}  \times U(1)_{f}  \times U(1)_{c} $ ( or equivalently
    $ SU(2)_{s}  \times U(1)_{c1}  \times U(1)_{c2}$, because we have two
   independent U(1) charge symmetries in the channel 1 and the channel 2 ).

  In this section, for simplicity, we take
  $ \lambda^{x}=\lambda^{y}=\lambda, \lambda^{z} \neq \lambda; 
 \alpha^{x}=\alpha^{y}=\alpha, \alpha^{z} \neq \alpha $.
  The symmetry in the spin sector is reduced to $ U(1) \times Z_{2} 
  \sim O(2) $ \cite{semi}.  In the following, we closely follow
   the notations of Emery-Kivelson \cite{emery}. Abelian-bosonizing
  the four bulk Dirac fermions separately:
\begin{equation}
 \psi_{i \alpha }(x )= \frac{P_{i \alpha}}{\sqrt{ 2 \pi a }}
  e ^{- i \Phi_{i \alpha}(x) }
\label{first}
\end{equation}
    Where  $ \Phi_{i \alpha} (x) $ are the real chiral bosons satisfying
 the commutation relations
\begin{equation}
   [ \Phi_{i \alpha} (x), \Phi_{j \beta} (y) ]
   =   \delta_{i j} \delta_{\alpha \beta} i \pi sgn( x-y )
\end{equation}
   
    The cocyle factors have been chosen as: $ P_{1 \uparrow}= P_{1 \downarrow}
  = e^{i \pi N_{1 \uparrow} }, P_{2 \uparrow}= P_{2 \downarrow}
  = e^{i \pi ( N_{1 \uparrow} + N_{1 \downarrow} + N_{2 \uparrow} ) } $.

   It is convenient to introduce the following charge, spin, flavor,
  spin-flavor bosons:
\begin{eqnarray}
 \Phi_{c} & = & \frac{1}{2} ( \Phi_{1 \uparrow }+ \Phi_{1 \downarrow }+
  \Phi_{2 \uparrow }+ \Phi_{2 \downarrow } )   \nonumber \\
 \Phi_{s} & = & \frac{1}{2} ( \Phi_{1 \uparrow }- \Phi_{1 \downarrow }+
  \Phi_{2 \uparrow }- \Phi_{2 \downarrow } )   \nonumber \\
 \Phi_{f} & = & \frac{1}{2} ( \Phi_{1 \uparrow }+ \Phi_{1 \downarrow }-
  \Phi_{2 \uparrow }- \Phi_{2 \downarrow } )   \nonumber \\
 \Phi_{sf}& = & \frac{1}{2} ( \Phi_{1 \uparrow }- \Phi_{1 \downarrow }-
  \Phi_{2 \uparrow }+ \Phi_{2 \downarrow } )
\label{second}  
\end{eqnarray}

    The spin currents $ J^{a}(x) = J^{a}_{1}(x) + J^{a}_{2}(x) $
  and $ \tilde{J}^{a}(x) = J^{a}_{1}(x) - J^{a}_{2}(x) $ can be expressed
  in terms of the above chiral bosons
\begin{eqnarray}
  J_{x}= \frac{1}{\pi a} \cos \Phi_{s} \cos \Phi_{sf},~
  J_{y}= \frac{1}{\pi a} \sin \Phi_{s} \cos \Phi_{sf},~
  J_{z}= -\frac{1}{ 2 \pi} \frac{ \partial \Phi_{s}}{\partial x}  \nonumber  \\
  \tilde{J}_{x}=- \frac{1}{\pi a} \sin \Phi_{s} \sin \Phi_{sf},~
  \tilde{J}_{y}= \frac{1}{\pi a} \cos \Phi_{s} \sin \Phi_{sf},~
  \tilde{J}_{z}= -\frac{1}{ 2 \pi} \frac{ \partial \Phi_{sf}}{\partial x}
\label{current}
\end{eqnarray}
  
   After making the canonical transformation $ U= \exp [ i S^{z} \Phi_{s}(0)] $
   and the following refermionization
\begin{eqnarray}
S^{x} &= & \frac{ \widehat{a}}{\sqrt{2}} e^{i \pi N_{sf}},~~~
S^{y}= \frac{ \widehat{b}}{\sqrt{2}} e^{i \pi N_{sf}},~~~
S^{z}= -i \widehat{a} \widehat{b}        \nonumber \\
 \psi_{sf} & = & \frac{1}{\sqrt{2}}( a_{sf} - i b_{sf} ) =
  \frac{1}{\sqrt{ 2 \pi a}} e^{i \pi N_{sf}} e^{-i \Phi_{sf} }     \nonumber   \\
 \psi_{s,i} & = & \frac{1}{\sqrt{2}}(  a_{s,i} - i  b_{s,i} )=
 \frac{1}{\sqrt{ 2 \pi a}} e^{i \pi( d^{\dagger}d + N_{sf})} e^{-i \Phi_{s} }   
\label{refer}
\end{eqnarray}

  The transformed Hamiltonian $ H^{\prime}= U H U^{-1} =
    H_{sf} + H_{s} + \delta H $ can be written in terms of
    the Majorana fermions  \cite{atten}: 
\begin{eqnarray}
 H_{sf} &= & \frac{ i v_{F} }{2} \int dx (a_{sf}(x) \frac{ \partial a_{sf}(x)}
 {\partial x} + b_{sf}(x) \frac{ \partial b_{sf}(x)} {\partial x} )
   -i \frac{ \lambda }{\sqrt{ 2 \pi a}} \widehat{a} b_{sf}(0)
   +i \frac{ \alpha }{\sqrt{ 2 \pi a}} \widehat{b} a_{sf}(0)
                                                 \nonumber \\
 H_{s}& = & \frac{ i v_{F} }{2} \int dx (a_{s}(x) \frac{ \partial a_{s}(x)}
 {\partial x} + b_{s}(x) \frac{ \partial b_{s}(x)} {\partial x} )
        -i h \int dx a_{s}(x) b_{s}(x)   \nonumber  \\
  \delta H &= & -\lambda_{z}^{\prime} \widehat{a} \widehat{ b} a_{s}(0)
        b_{s}(0)
   -\alpha_{z} \widehat{a} \widehat{b} a_{sf}(0) b_{sf}(0)
\label{anderson}
\end{eqnarray}
  where $ \lambda_{z}^{\prime} = \lambda^{z} - 2 \pi v_{F} $. 

   It is instructive to compare the above equation with Eq.3 in 
 Ref.\cite{sf}. They looks very similar:
 {\em half} of the impurity spin coupled to half of the
 spin-flavor electrons, {\em another half} of the impurity spin coupled
 to {\em another} half of the spin-flavor electrons.
 However {\em the two canonical transformations employed in the two models
  are different}.
 This fact make the boundary conditions of this model rather different
 from that of the two channel spin-flavor Kondo model (2CSFK) discussed in Ref.\cite{sf}.

   The above Hamiltonian was first derived by Ref.\cite{gogolin}
 using Anderson-Yuval's approach. They found the solvable line $
 \lambda^{z} =2 \pi v_{F}, \alpha^{z}=0 $ and calculated the exact crossover
 function of free energy
 along this solvable line. Using EK's method, We rederived
 this Hamiltonian \cite{trivial}.  The huge advantage of EK's method over Anderson-Yuval's
 approach is that the {\em boundary conditions} at different boundary 
 fixed points can be identified \cite{powerful}.

 By using the Operator Product Expansion (OPE) of the various
  operators in Eq.\ref{anderson} \cite{cardy},
    we get the RG flow equations near the weak coupling fixed
    point $\lambda_{z}=2 \pi v_{F}, \lambda=\alpha=\alpha_{z}=0 $
\begin{eqnarray}
  \frac{ d \lambda}{d l} & = &\frac{1}{2} \lambda+ \alpha \alpha_{z}
                       \nonumber   \\
  \frac{ d \alpha}{d l} & = &\frac{1}{2} \alpha -
                   \lambda \alpha_{z}   \nonumber  \\
  \frac{ d \alpha_{z}}{d l} & = & -\lambda \alpha
\label{danger}
\end{eqnarray}
 
     The fact that we find {\em two} relevant operators in the above equations
  may indicate there are {\em two} intermediate coupling fixed points.
  However, in the following, the two intermediate coupling fixed points are
  shown to be the same.

   The {\em original} impurity spin in $ H $ are related to those in $ H^{\prime} $ by
\begin{eqnarray}
  S^{H}_{x}  &= & U S_{x} U^{-1} = S_{x} \cos \Phi_{s}(0) - S_{y} \sin \Phi_{s}(0)  \nonumber  \\
  S^{H}_{y} &= & U S_{y} U^{-1} = S_{x} \sin \Phi_{s}(0) + S_{y} \cos \Phi_{s}(0)  \nonumber  \\
  S^{H}_{z} &= & U S_{z} U^{-1} = S_{z}
\label{change}
\end{eqnarray}

   Using the refermionization Eq.\ref{refer}, the {\em original} impurity spin in $ H $ can
  be written in terms of fermions
\begin{eqnarray}
  S^{H}_{x}  &= & i( \widehat{b} a_{s,i}+\widehat{a} b_{s,i} )             \nonumber  \\
  S^{H}_{y}  &= & i( \widehat{b} b_{s,i}-\widehat{b} a_{s,i} )             \nonumber  \\
  S^{H}_{z} &= & -i \widehat{a} \widehat{b}
\label{imp}
\end{eqnarray}

  At $ \lambda^{\prime}_{z}=0 $, the spin boson $ \Phi_{s} $ completely decouples
  from the impurity in $ H^{\prime} $, therefore $ \chi_{imp} =0 $.
  Because the canonical transformation $ U $ is a boundary
  condition changing operator \cite{boundary,powerful}, at $ \lambda^{\prime}_{z} =0 $,
  this leads to
\begin{equation}
  a^{s}_{L}(0)=-a^{s}_{R}(0), ~~ b^{s}_{L}(0)=-b^{s}_{R}(0)
\label{bound1}
\end{equation}

  Following Ref.\cite{powerful}, in order to identify the fixed points along the solvable line 
 $ \lambda^{\prime}_{z}=0, \alpha_{z}=0 $ (we also set $ h=0 $),
 we write $ H_{sf} $ in the action form
\begin{eqnarray}
 S &= & S_{0} + \frac{\gamma_{1}}{2} \int d \tau \widehat{a}(\tau)
       \frac{\partial \widehat{a}(\tau)}{\partial \tau}
    + \frac{\gamma_{2}}{2} \int d \tau \widehat{b}(\tau)
       \frac{\partial \widehat{b}(\tau)}{\partial \tau}
                                 \nonumber   \\
   & - & i \frac{ \lambda }{\sqrt{ 2 \pi a}}
      \int d \tau \widehat{a}(\tau) b_{sf}(0, \tau)
      +i \frac{ \alpha }{\sqrt{ 2 \pi a}}
        \int d \tau \widehat{b}(\tau) a_{sf}(0,\tau)
\label{action}
\end{eqnarray}

  When performing the RG analysis of the action $ S $, we keep \cite{above}
   1: $ \gamma_{2}=1, \lambda $ fixed,
   2: $ \gamma_{1}=1, \alpha $ fixed,
   3: $ \lambda, \alpha $ fixed;
   three fixed points of Eq.\ref{anderson} can be identified

\subsection{ Fixed point 1}

  This fixed point is located at $ \gamma_{1}=0, \gamma_{2}=1 $
 where $ \widehat{b} $ decouples, but $ \widehat{a} $ loses its
 kinetic energy and becomes a Grassmann Lagrangian multiplier.
 Integrating $\widehat{a} $ out leads to
 the following boundary conditions \cite{trick}:
\begin{equation}
  b^{sf}_{L}(0)=-b^{sf}_{R}(0)
\label{bound2}
\end{equation}

  Eqs.\ref{bound1},\ref{bound2} can be expressed in terms of bosons:
\begin{equation}
\Phi_{s,L}(0)=\Phi_{s,R}(0)+\pi, ~~~ \Phi_{sf,L}(0)=-\Phi_{sf,R}(0)+\pi
\end{equation}

 This is just the non-fermi liquid fixed point of the 2CK.
 The three Majorana fermions in the spin sector being twisted, this fixed point
 possesses the symmetry $ O(3) \times O(5) $. The finite size spectrum
 of this fixed point was listed in Ref.\cite{powerful}.

  The local correlation functions at the 2CK fixed point are \cite{powerful}:
\begin{equation}
\langle \widehat{a}( \tau ) \widehat{a}(0) \rangle =\frac{1}{\tau},~~~~
\langle b_{sf}( \tau ) b_{sf}(0) \rangle =\frac{\gamma^{2}_{1}}{\tau^{3}}
\label{dimension}
\end{equation}

  From the above equation, we can read the scaling dimensions of the various
fields $ [\widehat{b}]=0, [\widehat{a}]=[a_{s}]=[b_{s}]=[a_{sf}]=1/2, [b_{sf}]=3/2 $.

 As shown in Ref.\cite{powerful}, at the fixed point, the impurity degree of freedoms
 completely disappear: $\widehat{b} $ decouples and $ \widehat{a} $ turns into the
 {\em non-interacting } scaling field at the fixed point \cite{care}
\begin{equation}
    \widehat{a} \sim  b_{sf}(0,\tau)
\end{equation}

   Using Eq.\ref{imp}, the impurity spin turns into
\begin{eqnarray}
  S^{H}_{x}(\tau)  &= & i( \widehat{b} a_{s,i}(0,\tau)+b_{sf}(0,\tau) b_{s,i}(0,\tau) )            \nonumber  \\
  S^{H}_{y}(\tau)  &= & i( \widehat{b} b_{s,i}(0,\tau)-b_{sf}(0,\tau) a_{s,i}(0,\tau) )             \nonumber  \\
  S^{H}_{z}(\tau) &= & i \widehat{b} b_{sf}(0,\tau)
\end{eqnarray}

   Using the relation
\begin{equation}
   \psi^{H}_{s}(x)= U \psi_{s}(x) U^{-1}=i sgnx \psi_{s,i}(x)
\label{reverse}
\end{equation}

   We get \cite{cut} 
\begin{eqnarray}
  S_{x}(\tau)  &= & i( -\widehat{b} b_{s}(0,\tau)+b_{sf}(0,\tau) a_{s}(0,\tau) )            \nonumber  \\
  S_{y}(\tau)  &= & i( \widehat{b} a_{s}(0,\tau)+b_{sf}(0,\tau) b_{s}(0,\tau) )             \nonumber  \\
  S_{z}(\tau) &= & i ( \widehat{b} b_{sf}(0,\tau) + a_{s}(0,\tau) b_{s}(0,\tau) )
\label{add}
\end{eqnarray}

The impurity spin-spin correlation function $ \langle S^{a}(\tau) S^{a}(0) \rangle
  =\frac{1}{\tau} $.

   The above equations \cite{add} are consistent with the CFT identifications \cite{line}
\begin{equation}
  \vec{S} \sim \vec{\phi} + \vec{ J} + \cdots
\end{equation}

 The 2CK fixed point is unstable, because there is a dimension 1/2 relevant
 operator $ \widehat{b} a_{sf} $, the OPE of $ a_{sf} $ with itself will
 generate the dimension 2 energy momentum tensor of this Majorana fermion
 $  T(\tau)= \frac{1}{2} a_{sf}(0,\tau) \frac{ \partial a_{sf}(0,\tau)}{\partial \tau} $,
 The OPE of the energy momentum tensor with the primary field $ a_{sf} $ is
\begin{equation}
   T(\tau_{1}) a_{sf}(\tau_{2})=  \frac{ \frac{1}{2} a_{sf}(\tau_{2})}{ (\tau_{1}-\tau_{2})^{2}}
    + \frac{ L_{-1} a_{sf}(\tau_{2})}{\tau_{1}-\tau_{2}} + L_{-2} a_{sf}(\tau_{2}) + \cdots
\end{equation}

 First order descendant field of this primary field $ L_{-1} a_{sf}(0,\tau)=
 \frac{ \partial a_{sf}(0,\tau)}{\partial \tau} $ with dimension 3/2 is generated.
 $ \lambda^{\prime}_{z} $ term in $ \delta H $ has scaling 
 dimension 3/2, it will generate a dimension 2 operator
 $ a_{s}(0,\tau) \frac{ \partial a_{s}(0,\tau)}{\partial \tau} +
 b_{s}(0,\tau) \frac{ \partial b_{s}(0,\tau)}{\partial \tau} $. $\gamma_{2} $ term
 has dimension 2 also.

 From Eq.\ref{dimension}, we can see $ \alpha_{z} $ term has scaling dimension 5/2, 
it can be written as
\begin{equation}
    :\widehat{a}(\tau) \frac{\partial \widehat{a}(\tau)}{\partial \tau}: a_{sf}(0,\tau)
 = :b_{sf}(0,\tau) \frac{\partial b_{sf}(0,\tau)}{\partial \tau}: a_{sf}(0,\tau)
\end{equation}
  
     The bosonized form of this operator is
\begin{equation}
  :( \cos 2\Phi_{sf}(0,\tau)-\frac{1}{2} (\partial \Phi_{sf}(0,\tau))^{2}): \sin \Phi_{sf}(0,\tau)
\end{equation}

 Using CFT, Ref.\cite{affleck} predicted a dimension 1/2 relevant operator
 $ \phi^{3}_{f} $ in the flavor sector. Ref.\cite{line} classified all the
 first order descendants of the primary operator in the spin sector.
 In the flavor sector, the same classification apply,
 $ \vec{J}_{-1} \cdot \vec{\phi}_{f} $ is Charge-Time Reversal (CT) odd,
 therefore is not allowed,
 but $ L_{-1} \phi^{3}_{f} $ is CT even.  The CFT analysis is completely
 consistent with the above EK's solution.

  In order to make this fixed point stable, we have to tune $ \alpha
  =\alpha_{z}=0 $, namely the channel anisotropy is strictly prohibited.
  If $\alpha=0 $, but $ \alpha_{z} \neq 0 $, because
  $ \alpha_{z} $ is highly irrelevant, it {\em seems} the 2CK
  fixed point is stable. However, this is not true. From the RG flow Eq.\ref{danger}, it
 is easy to see that even initialy $ \alpha=0 $, it will be generated,
  $ \alpha_{z} $ is 'dangerously' irrelevant. 

\subsection{ Fixed point 2}

  This fixed point is located at $ \gamma_{1}=1, \gamma_{2}=0 $
 where $ \widehat{a} $ decouples, but $ \widehat{b} $ loses its
 kinetic energy and becomes a Grassmann Lagrangian multiplier.
 Integrating $\widehat{b} $ out leads to
 the following boundary conditions:
\begin{equation}
  a^{sf}_{L}(0)=-a^{sf}_{R}(0)
\label{dual}
\end{equation}

  Eqs.\ref{bound1},\ref{dual} can be expressed in terms of bosons:
\begin{equation}
\Phi_{s,L}(0)=\Phi_{s,R}(0)+\pi, ~~~ \Phi_{sf,L}(0)=-\Phi_{sf,R}(0)
\end{equation}
    
    This fixed point also possesses the symmetry $ O(3) \times O(5) $.
  In fixed points 1 and 2, $\widehat{a} $ and
  $\widehat{b} $, $ b_{sf} $ and $ a_{sf} $ exchange roles.

  As discussed in the fixed point 1, $ \alpha_{z} $ is 'dangerously' irrelevant.
   In order to make this fixed point stable, we have to tune $ \lambda
   =\alpha_{z}=0 $.  This fixed point
   is actually {\em the same} with the 2CK fixed point. This can be seen most
   clearly from the original Eq.\ref{kondob}: if $\lambda=\alpha_{z}
    =0 $, under the $ SU(2) $ transformation on the channel 2 fermions
   $ \psi_{2 \uparrow} \rightarrow i \psi_{2 \uparrow}, 
    \psi_{2 \uparrow} \rightarrow -i \psi_{2 \uparrow} $, the spin currents
   of channel 2 transform as
    $ J^{x}_{2} \rightarrow -J^{x}_{2}, J^{y}_{2} \rightarrow -J^{y}_{2},
      J^{z}_{2} \rightarrow J^{z}_{2} $, Eq.\ref{kondob} is transformed back to
    the 2 channel flavor symmetric Kondo model. This can also be seen from
    Eq.\ref{current}, $ \tilde{J}_{x}, \tilde{J}_{y}, J_{z} $ also satisfy
    the $ \widehat{SU}_{2}(2) $ algebra.
 
\subsection{ Fixed point 3}

 This fixed point is located at $ \gamma_{1}=\gamma_{2}=0 $
 where both $\widehat{a} $ and $ \widehat{b} $
 lose their kinetic energies and become
 two Grassmann Lagrangian multipliers. Integrating them out leads to
 the following boundary conditions:
\begin{equation}
  b^{sf}_{L}(0)=-b^{sf}_{R}(0), ~~ a^{sf}_{L}(0)=-a^{sf}_{R}(0)
\label{bound3}
\end{equation}

  Eqs.\ref{bound1}, \ref{bound3} can be expressed in term of bosons:
\begin{equation}
   \Phi^{s}_{L}=\Phi^{s}_{R} + \pi, ~~~ \Phi^{sf}_{L}=\Phi^{sf}_{R} + \pi 
\end{equation}

   Substituting the above equation to Eqs. \ref{first} \ref{second} and paying
   attention to the {\em spinor} nature of the representation \cite{hopping},
   it is easy to see that depending on the sign of $\alpha$,
  {\em one} of the two channels suffer $\frac{\pi}{2} $ phase shift,
  {\em the other} remains free. The four Majorana fermions being twisted,
   this fixed point has the symmetry $ O(4) \times O(4) $ with $ g=1 $.
   The finite size spectrum of this fixed point is listed in Table \ref{flavor},
   it is the sum of 
   that with phase shift $ \pi/2 $ and that of free electrons.
   This scenario is completely consistent with NRG results of Ref.\cite{cox2}.

  The local correlation functions at the FL fixed point are \cite{powerful}:
\begin{eqnarray}
\langle \widehat{a}( \tau ) \widehat{a}(0) \rangle =\frac{1}{\tau},~~~~
\langle b_{sf}( \tau ) b_{sf}(0) \rangle =\frac{\gamma^{2}_{1}}{\tau^{3}}  \nonumber  \\
\langle \widehat{b}( \tau ) \widehat{b}(0) \rangle =\frac{1}{\tau},~~~~
\langle a_{sf}( \tau ) a_{sf}(0) \rangle =\frac{\gamma^{2}_{2}}{\tau^{3}}  
\end{eqnarray}

  From the above equation, We can read the scaling dimensions of the various fields:
  $[\widehat{a}]=[\widehat{b}]=[a_{s}]=[b_{s}]=1/2, [a_{sf}]=[b_{sf}]=3/2 $.

 At the fixed point, the impurity degree of freedoms
 completely disappear: $\widehat{a}, \widehat{b} $ turn into the
 {\em non-interacting } scaling fields at the fixed point
\begin{equation}
    \widehat{a} \sim  b_{sf}(0,\tau),~~~ \widehat{b} \sim  a_{sf}(0,\tau)
\end{equation}

   Using Eqs.\ref{imp}, \ref{reverse}, the impurity spin turns into
\begin{eqnarray}
  S_{x}(\tau)  &= & i( -a_{sf}(0,\tau) b_{s}(0,\tau)+b_{sf}(0,\tau) a_{s}(0,\tau) )            \nonumber  \\
  S_{y}(\tau)  &= & i( a_{sf}(0,\tau) a_{s}(0,\tau)+b_{sf}(0,\tau) b_{s}(0,\tau) )             \nonumber  \\
  S_{z}(\tau) &= & i ( a_{sf}(0,\tau) b_{sf}(0,\tau) + a_{s}(0,\tau) b_{s}(0,\tau) )
\end{eqnarray}

The impurity spin-spin correlation function show typical FL behavior
\begin{equation}
  \langle S^{z}(\tau) S^{z}(0) \rangle =\frac{1}{\tau^{2}} 
\end{equation}

   Using the fermionized form of the Eq.\ref{current} and paying attention to
  the {\em spinor} nature of the representation,
  it is easy to see the impurity spin renormalizs into either $ \vec{J}_{1} (0,\tau) $
  or $ \vec{J}_{2}(0,\tau) $ depending on the sign of $\alpha $. This is consistent
  with the CFT analysis in the Appendix. 

  There are 4 leading irrelevant operators with dimension 2 in
  the action $ S $ : $ \gamma_{1} $ and $ \gamma_{2} $ terms,
  $\lambda_{z}^{\prime} $ term and
 $ a_{s}(0,\tau) \frac{ \partial a_{s}(0,\tau)}{\partial \tau} +
  b_{s}(0,\tau) \frac{ \partial b_{s}(0,\tau)}{\partial \tau} $ which
  will be generated by the $\lambda_{z}^{\prime} $ term. 

  The $ \alpha_{z} $ term has dimension 4, it can be written as $
    :\widehat{a}(\tau) \frac{\partial \widehat{a}(\tau)}{\partial \tau}:
    :\widehat{b}(\tau) \frac{\partial \widehat{b}(\tau)}{\partial \tau}: $.

  The bosonized forms of the 4 leading irrelevant operators are \cite{another}
\begin{eqnarray}
    \widehat{a}(\tau) \frac{\partial \widehat{a}(\tau)}{\partial \tau}
 & = & \cos 2\Phi_{sf}-\frac{1}{2} (\partial \Phi_{sf}(0))^{2}  \nonumber   \\
    \widehat{b}(\tau) \frac{\partial \widehat{b}(\tau)}{\partial \tau}
 & = & -\cos 2\Phi_{sf}-\frac{1}{2} (\partial \Phi_{sf}(0))^{2}  \nonumber   \\
\widehat{a}\widehat{b}a_{s}(0)b_{s}(0) & = &   \partial\Phi_{sf}(0,\tau)
  \partial\Phi_{s}(0,\tau)   \nonumber   \\
    a_{s}(0,\tau) \frac{ \partial a_{s}(0,\tau)}{\partial \tau} & + &
  b_{s}(0,\tau) \frac{ \partial b_{s}(0,\tau)}{\partial \tau}
  = (\partial \Phi_{s}(0,\tau))^{2}    
\label{four}
\end{eqnarray}

  Following the method developed in Ref.\cite{powerful}, their contributions
 to the single particle Green functions can be calculated.
 The first order correction
  to the single particle L-R Green function ( $ x_{1}>0, x_{2}<0 $ ) 
  due to the first operator in the above Eq. is
\begin{eqnarray}
  &\langle & \psi_{1 \uparrow}( x_{1},\tau_{1} ) \psi^{\dagger}_{1 \uparrow}( x_{2},\tau_{2} ) \rangle  =
  \int d\tau
 \langle e^{-\frac{i}{2} \Phi_{c}( x_{1}, \tau_{1} )} e^{\frac{i}{2} \Phi_{c}( x_{2}, \tau_{2} )}\rangle 
                        \nonumber   \\  
 & \times &\langle  e^{-\frac{i}{2} \Phi_{s}( x_{1}, \tau_{1} )}
 e^{\frac{i}{2} ( \Phi_{s}( x_{2}, \tau_{2} ) + \pi )} \rangle  
  \langle  e^{-\frac{i}{2} \Phi_{f}( x_{1}, \tau_{1} )} e^{\frac{i}{2} \Phi_{f}( x_{2}, \tau_{2} )}\rangle
                          \nonumber   \\
 & \times & \langle  e^{-\frac{i}{2} \Phi_{sf}( x_{1}, \tau_{1} )}
 (:\cos2 \Phi_{sf}( 0, \tau ): -\frac{1}{2} : (\partial \Phi_{sf}( 0,\tau) )^{2} :) 
 e^{\frac{i}{2} ( \Phi_{sf}( x_{2}, \tau_{2} ) +\pi )}\rangle   \nonumber  \\
  & \sim &  (z_{1}-\bar{z}_{2} )^{-2}
\label{single}
\end{eqnarray}
  Where $ z_{1}=\tau_{1}+i x_{1} $ is in the upper half plane,
  $ \bar{z}_{2}  =\tau_{2}+i x_{2} $ is in the lower half plane.

  By using the following OPE:
\begin{eqnarray}
 : e^{-\frac{i}{2} \Phi_{sf}( z_{1} )}: : e^{\frac{i}{2}  \Phi_{sf}( z_{2})}: =
 (z_{1}-z_{2})^{-1/4}-\frac{i}{2}(z_{1}-z_{2})^{3/4} :\partial \Phi_{sf}(z_{2}):  
                            \nonumber   \\
 -\frac{i}{4}(z_{1}-z_{2})^{7/4} :\partial^{2} \Phi_{sf}(z_{2}):
 -\frac{1}{8}(z_{1}-z_{2})^{7/4} : (\partial \Phi_{sf}(z_{2}) )^{2}: + \cdots
\label{ope}
\end{eqnarray}

   It is ease to see that the primary field $ :\cos2 \Phi_{sf}( 0, \tau ): $ makes {\em no} contributions
to the three point function.
 It was shown by the detailed calculations
 in Ref.\cite{conductivity} that only the part of the self-energy which is both {\em imaginary}
 and {\em even} function of $ \omega $ contributes to the conductivity.
 Although the energy momentum tensor $ : (\partial \Phi_{sf}( 0,\tau) )^{2} : $
do make $\sim \omega $  contribution to the self-energy in the first order \cite{conn},
 because it is a {\em odd} function, it does {\em not} contribute to the electron conductivity in this order.
 Same arguments apply to the other operators in Eq.\ref{four}.
 Second order perturbations in these operators lead to the generic $ T^{2} $ fermi
 liquid bahaviour of the electron conductivity.

   The results of this section were applied to a two level tunneling system with slight
  modifications in Ref.\cite{hopping}. The universal scaling functions in the presence
 of external magnetic field which breaks the channel symmetry were also discussed there.

\section{Compactified one channel Kondo Model}

   Assuming Particle-Hole symmetry, the {\em non-interacting} one channel
 Kondo model has two commuting $ SU(2) $ symmetry,
 one is the usual spin symmetry with the generators $ J^{a} (a=x,y,z) $
 another is the isospin symmetry with the generators $ I^{a} (a=x,y,z) $.
\begin{eqnarray}
J_{x} & =  & \frac{1}{2}( \psi^{\dagger}_{\uparrow} \psi_{\downarrow}
    + \psi^{\dagger}_{\downarrow} \psi_{\uparrow} ), ~~
J_{y}=\frac{1}{2i}( \psi^{\dagger}_{\uparrow} \psi_{\downarrow}
    - \psi^{\dagger}_{\downarrow} \psi_{\uparrow} ), ~~
J_{z}=\frac{1}{2}( \psi^{\dagger}_{\uparrow} \psi_{\uparrow}
    - \psi^{\dagger}_{\downarrow} \psi_{\downarrow} )   \nonumber   \\
I_{x} & =  & \frac{1}{2}( \psi^{\dagger}_{\uparrow} \psi^{\dagger}_{\downarrow}
    + \psi_{\downarrow} \psi_{\uparrow} ), ~~
I_{y}=\frac{1}{2i}( \psi^{\dagger}_{\uparrow} \psi^{\dagger}_{\downarrow}
    - \psi_{\downarrow} \psi_{\uparrow}), ~~
I_{z}=\frac{1}{2}( \psi^{\dagger}_{\uparrow} \psi_{\uparrow}
    + \psi^{\dagger}_{\downarrow} \psi_{\downarrow} )
\label{si}
\end{eqnarray}

   The diagonal and off-diagonal components of the isospin currents represent respectively
 the charge and pairing density at  the site $ x$.

  The one channel compactified model proposed by Ref.\cite{coleman} is a model
 where the impurity spin couples to both the spin and the isospin currents of the
 one channel conduction electrons
 \begin{eqnarray}
 H_{c} &= & i v_{F} \int^{\infty}_{-\infty} dx 
   \psi^{\dagger}_{ \alpha }(x) \frac{d \psi_{ \alpha }(x)}{dx}
   + \sum_{a=x,y,z} \lambda^{a}( I^{a}(0)+ J^{a}(0) )  S^{a} 
   + \sum_{a=x,y,z} \alpha^{a}  
    (I^{a}(0)-J^{a}(0)) S^{a}       \nonumber \\
  & + & h ( \int dx (I^{z}(x) + J^{z}(x)) + S^{z} )
\label{com}
 \end{eqnarray}

   The ordinary symmetric Anderson impurity model in a one dimensional lattice is
\begin{eqnarray}
  H & = & i t \sum_{n,\alpha} ( \psi^{\dagger}_{\alpha}(n+1) \psi_{\alpha}(n) - h. c. )  \nonumber  \\
     & + & i V \sum_{\alpha} ( \psi^{\dagger}_{\alpha}(0) d_{\alpha} -h.c.)
     + U(n_{d \uparrow}-\frac{1}{2})(n_{d \downarrow}-\frac{1}{2})
\label{aim}
\end{eqnarray}

    The $ O(4) $ symmetry of the AIM can be clearly displayed in terms of the
    Majorana fermions
\begin{eqnarray}
 \psi_{\uparrow}(n) & = &\frac{1}{\sqrt{2}} ( \chi_{1}(n) -i \chi_{2}(n) ), ~~~
  d_{\uparrow}=\frac{1}{\sqrt{2}}( d_{1}-i d_{2} )    \nonumber   \\
 \psi_{\downarrow}(n) & = & \frac{1}{\sqrt{2}} ( -\chi_{3}(n) -i \chi_{0}(n) ), ~~~
  d_{\downarrow}=\frac{1}{\sqrt{2}}( -d_{3}-i d_{0} )
\end{eqnarray}

    Breaking the symmetry from $ O(4) $ to $ O(3) \times O(1) $ in the hybridization \cite{ising},
 the Hamiltonian \ref{aim} becomes:
\begin{eqnarray}
 H & = & i t \sum_{n} \sum^{3}_{\alpha=0} \chi_{\alpha}(n+1) \chi_{\alpha}(n) +i V_{0} \chi_{0}(0) d_{0}
                                              \nonumber   \\
  & + & i V \sum^{3}_{ \alpha=1} \chi_{\alpha}(0) d_{\alpha} + U d_{1} d_{2} d_{3} d_{0}
\label{break}
\end{eqnarray}

     In the large $ U $ limit, projecting out the excited impurity states, we can map
  the  Hamiltonian \ref{break} to the 1CCK Hamiltonian \ref{com} with
\begin{equation}
  \lambda= \frac{ 2 V^{2}}{U},~~~ \alpha= -\frac{ 2 V_{0} V }{ U }
\end{equation}

     If $ V_{0}=V $, Eq.\ref{break} comes back to the original $ O(4) $ symmetric AIM. In the strong
 coupling limit, it becomes the one channel Kondo model where the impurity only couples to
 the spin currents (or isospin currents) of the conduction electrons \cite{exchange}.

 If $ V_{0} =0 $, then $ \alpha=0 $, Eq.\ref{break} becomes the isotropic 1CCK where the impurity couples to
 the spin and isospin currents with equal strength.
 If we define the P-H transformation $ \psi_{\uparrow} \rightarrow \psi_{\uparrow},
 \psi_{\downarrow} \rightarrow \psi^{\dagger}_{\downarrow} $, then spin and isospin currents
 transform to each other $ I^{a} \rightarrow J^{a}, J^{a} \rightarrow I^{a} $.
  The Hamiltonian \ref{com} has the P-H symmetry if $\alpha=0 $.

   In the following, parallel to the discussions on the 2CFAK, we take
  $ \lambda^{x}=\lambda^{y}=\lambda, \lambda^{z} \neq \lambda; 
 \alpha^{x}=\alpha^{y}=\alpha, \alpha^{z} \neq \alpha $.
    We bosonize the spin $\uparrow $ and spin $\downarrow $ electrons separately
\begin{equation}
 \psi_{ \alpha }(x )= \frac{P_{ \alpha}}{\sqrt{ 2 \pi a }}
  e ^{- i \Phi_{ \alpha}(x) }
\label{one}
\end{equation}
   
    The cocyle factors have been chosen as $ P_{ \uparrow}= P_{ \downarrow}
  = e^{i \pi N_{ \uparrow} } $.

 The bosonized form of the spin and isospin currents in Eq.\ref{si}  are
\begin{eqnarray}
  J_{x} & = & \frac{1}{2\pi a} \cos \sqrt{2} \Phi_{s},~~~
  J_{y}=\frac{1}{2\pi a} \sin \sqrt{2} \Phi_{s},~~~
  J_{z}=-\frac{1}{4\pi } \frac{\partial}{\partial x} \sqrt{2} \Phi_{s}  \nonumber  \\
  I_{x} & = & \frac{1}{2\pi a} \cos \sqrt{2} \Phi_{c},~~~
  I_{y}=\frac{1}{2\pi a} \sin \sqrt{2} \Phi_{c},~~~
  I_{z}=-\frac{1}{4\pi } \frac{\partial}{\partial x} \sqrt{2} \Phi_{c}
\end{eqnarray}
   where $ \Phi_{c}, \Phi_{s} $ are charge and spin bosons:
\begin{eqnarray}
  \Phi_{c}  = \frac{1}{\sqrt{2}}( \Phi_{\uparrow}+\Phi_{\downarrow}),~~~~~
  \Phi_{s}  =  \frac{1}{\sqrt{2}}( \Phi_{\uparrow}-\Phi_{\downarrow})
\label{cs}
\end{eqnarray}

    The sum $ J_{s}^{a}(x) = I^{a}(x) + J^{a}(x) $
  and the difference $ J_{d}^{a}(x) = I^{a}(x) - J^{a}(x) $ can be expressed
  in terms of the chiral bosons
\begin{eqnarray}
  J^{s}_{x}= \frac{1}{\pi a} \cos \Phi_{\uparrow} \cos \Phi_{\downarrow},~
  J^{s}_{y}= \frac{1}{\pi a} \sin \Phi_{\uparrow} \cos \Phi_{\downarrow},~
  J^{s}_{z}= -\frac{1}{ 2 \pi} \frac{ \partial \Phi_{\uparrow}}{\partial x}  \nonumber  \\
  J^{d}_{x}=- \frac{1}{\pi a} \sin \Phi_{\uparrow} \sin \Phi_{\downarrow},~
  J^{d}_{y}= \frac{1}{\pi a} \cos \Phi_{\uparrow} \sin \Phi_{\downarrow},~
  J^{d}_{z}= -\frac{1}{ 2 \pi} \frac{ \partial \Phi_{\downarrow}}{\partial x}
\label{iso}
\end{eqnarray}

    Compare Eq.\ref{current} with Eq.\ref{iso}, we immediately realize that
  the mapping between the 2CFAK and the 1CCK
  is $ \Phi_{s} \rightarrow \Phi_{\uparrow},  \Phi_{sf} \rightarrow \Phi_{\downarrow} $,
  therefore $ \psi_{s} \rightarrow \psi_{\uparrow},
  \psi_{sf} \rightarrow \psi_{\downarrow} $. The following fixed point structure
  can be immediately borrowed from the corresponding discussions on the 2CFAK.

\subsection{ Fixed point 1 }

   The boundary conditions are
\begin{equation}
  \psi_{\uparrow,L} = -\psi_{\uparrow,R},~~~
  \psi_{\downarrow,L} =\psi^{\dagger}_{\downarrow,R}
\end{equation}

   It is easy to see that the above boundary conditions respect the P-H symmetry,
   they can be expressed in terms of bosons
\begin{equation}
  \Phi_{\uparrow,L} =\Phi_{\uparrow,R} + \pi,~~~
  \Phi_{\downarrow,L} = -\Phi_{\downarrow,R} + \pi
\end{equation}

   Spin $\uparrow $ electrons suffer a $ \frac{\pi}{2} $ phase shift, however,
  spin $\downarrow $ electrons are scattered into holes and vice-versa.
 The one particle S-matrix are $ S_{\uparrow}=-1, S_{\downarrow}=0 $. The residual
 conductivity of the spin $\uparrow$ electron takes unitary limit, but that of the spin
 $\downarrow $ is half of the unitary limit. 
  The isotropic 1CCK has the same thermodynamic behaviors as the 2CK, but its 
  fixed point has the local KM symmetry $ \widehat{O}_{1}(3) \times \widehat{O}_{1}(1) $. 
  The finite size spectrum of this NFL fixed point is listed in
 Table \ref{compactnfl}. Comparing this finite size spectrum with that of the
 NFL fixed point of the 2CK listed in Ref.\cite{powerful}, it is easy to see that
 it has the {\em same } energy levels as those of the 2CK, but the corresponding
 degeneracy is {\em much smaller}. This is within the expectation, because
 the central charge $ c=2 $ and the fixed point symmetry of the isotropic 1CCK
  is smaller than that of the 2CK.

 This fixed point is stable only when $\alpha=\alpha_{z}=0 $ where
 the Hamiltonian \ref{com} has P-H symmetry.

   Away from the fixed point, there is only one dimension 3/2 operator
\begin{equation}
 \widehat{a} \widehat{b} \partial \Phi_{\uparrow}(0) \sim
   \cos \Phi_{\downarrow}(0) \partial \Phi_{\uparrow}(0) 
\end{equation}

  The first order correction to the single particle L-R Green function ( $ x_{1}>0, x_{2}<0 $ ) 
  due to this operator is
\begin{eqnarray}
  \int d\tau \langle e^{- i \Phi_{\uparrow}( x_{1}, \tau_{1} )} \partial \Phi_{\uparrow}(0,\tau)
     e^{ i ( \Phi_{\uparrow}( x_{2}, \tau_{2} ) +\pi )}\rangle 
 \langle :\cos \Phi_{\downarrow}( 0, \tau ): \rangle   =0    \nonumber   \\
  \int d\tau \langle e^{- i \Phi_{\downarrow}( x_{1}, \tau_{1} )} \cos \Phi_{\downarrow}(0,\tau)
     e^{- i \Phi_{\downarrow}( x_{2}, \tau_{2} )}\rangle 
 \langle \partial \Phi_{\uparrow}( 0, \tau ) \rangle   =0   
\end{eqnarray}

   By Wick theorem, it is easy to see that any {\em odd} order corrections vanish.

   Second order correction goes as $ \sim \omega $ which is a {\em odd} function, therefore
   does not contribute to the electron conductivity. The fourth order
   makes $ T^{2} $ contributions.

  There are two dimension 2 operators:
\begin{eqnarray}
  a_{\uparrow}(0,\tau) \frac{ \partial a_{\uparrow}(0,\tau)}{\partial \tau} & + &
  b_{\uparrow}(0,\tau) \frac{ \partial b_{\uparrow}(0,\tau)}{\partial \tau}   =   
       (\partial \Phi_{\uparrow}(0,\tau))^{2}     \nonumber   \\
  \widehat{a}(\tau) \frac{\partial \widehat{a}(\tau)}{\partial \tau}
 & = & \cos 2\Phi_{\downarrow}-\frac{1}{2} (\partial \Phi_{\downarrow}(0))^{2}  
\label{cool}
\end{eqnarray}

   The first order correction to the spin $\uparrow $ electron L-R Green function  
  due to the first operator in Eq.\ref{cool} is
\begin{eqnarray}
  \int d\tau \langle e^{- i \Phi_{\uparrow}( x_{1}, \tau_{1} )} :( \partial \Phi_{\uparrow}(0,\tau) )^{2}:
     e^{ i ( \Phi_{\uparrow}( x_{2}, \tau_{2} ) +\pi )}\rangle \sim ( z_{1}-\bar{z}_{2} )^{-2}
\label{many}
\end{eqnarray}

  As pointed out in the last section, the energy momentum tensor $ : (\partial \Phi_{\uparrow}( 0,\tau) )^{2} : $
 makes $\sim \omega $ contribution to the self-energy in the first order, therefore
 does not contribute to the electron conductivity.
 Second order perturbation in this operator leads to $ T^{2} $ contributions.

   Adding the contributions from all the leading irrelevant operators, we get
\begin{equation}
   \sigma_{\uparrow}(T) \sim \sigma_{u}(1+ T^{2} + T^{4} + \cdots)  
\end{equation}
  
   The first order correction to the spin $\downarrow $ electron L-R Green function  
  due to the 2nd operator in Eq. \ref{cool} is
\begin{eqnarray}
  \int d\tau \langle e^{- i \Phi_{\downarrow}( x_{1}, \tau_{1} )} \cos 2\Phi_{\downarrow}(0,\tau)
     e^{- i \Phi_{\downarrow}( x_{2}, \tau_{2} )}\rangle     \nonumber   \\
  -\frac{1}{2} \int d\tau \langle e^{- i \Phi_{\downarrow}( x_{1}, \tau_{1} )}
       ( \partial \Phi_{\downarrow}(0,\tau) )^{2}
     e^{- i \Phi_{\downarrow}( x_{2}, \tau_{2} )}\rangle    
\end{eqnarray}

  By using the following OPE:
\begin{equation}
 : e^{-i \Phi_{\downarrow}( z_{1} )}: : e^{-i  \Phi_{\downarrow}( z_{2})}: =
 (z_{1}-z_{2}): e^{-i 2 \Phi_{\downarrow}( z_{2} )}: 
 -i(z_{1}-z_{2})^{2} : e^{-i 2 \Phi_{\downarrow}( z_{2} )} \partial \Phi_{\downarrow}(z_{2}): +\cdots  
\end{equation}

   It is ease to see that the {\em second} integral vanishes, but the {\em first} becomes
\begin{equation}
\frac{1}{ (z_{1}-\bar{z}_{2} )^{-1} } \int d \tau \frac{1}{ (z_{1}-\tau)^{2} (\tau-\bar{z}_{2} )^{2} }
\sim (z_{1}-\bar{z}_{2} )^{-2} 
\end{equation}

    Putting $ \Delta=1 $ in Eq. (3.52) of Ref.\cite{conductivity}, we find the imaginary and
  real parts of self-energy go as $ Im \Sigma(\omega, T=0)=0, Re \Sigma(\omega, T=0)
  \sim \omega $, therefore the first order perturbation does not contribute to the spin $\downarrow $
 electron conductivity. Second order perturbation
 yields a $ T^{2} $ contributions.

   Adding the contributions from all the leading irrelevant operators, we get
\begin{equation}
  \sigma_{\downarrow}(T) \sim 2 \sigma_{u}(1+ T^{2}+ T^{4} + \cdots   )
\end{equation}

     The total conductivity is the summation of the two spin components \cite{bhatt}
\begin{equation}
  \sigma(T)= \sigma_{\uparrow}(T)+ \sigma_{\downarrow}(T)
   \sim 3 \sigma_{u}(1+ T^{2}+ T^{4} + \cdots   )
\end{equation}

  The boundary OPE of the spin and density of the {\em conduction electrons} are
\begin{eqnarray}
\psi^{\dagger}_{\uparrow}(z_{1}) \psi_{\uparrow}( \bar{z}_{2} ) & = & 
    ( z_{1}-\bar{z}_{2} )^{-1} +i \partial \Phi_{\uparrow} + \cdots     \nonumber   \\
\psi^{\dagger}_{\downarrow}(z_{1}) \psi_{\downarrow}( \bar{z}_{2} ) & = &
       0 + \cdots   \nonumber   \\ 
\psi^{\dagger}_{\uparrow}(z_{1}) \psi_{\downarrow}( \bar{z}_{2} ) & = & e^{i \sqrt{2} \Phi_{c}(0) } 
    + \cdots       \nonumber   \\
\psi^{\dagger}_{\downarrow}(z_{1}) \psi_{\uparrow}( \bar{z}_{2} ) & = &  e^{-i \sqrt{2} \Phi_{s}(0) } 
    + \cdots         
\label{sdnfl}
\end{eqnarray}

  The boundary OPE of the spin singlet and triplet pairing operators are
\begin{eqnarray}
\psi_{\uparrow}(z_{1}) \psi_{\uparrow}( \bar{z}_{2} ) & = & 0
    + \cdots       \nonumber   \\
\psi_{\downarrow}(z_{1}) \psi_{\downarrow}( \bar{z}_{2} ) & = & ( z_{1}-\bar{z}_{2} )^{-1}
    -i \partial \Phi_{\downarrow} + \cdots     \nonumber  \\
\psi_{\uparrow}(z_{1}) \psi_{\downarrow}( \bar{z}_{2} ) & = & e^{-i \sqrt{2} \Phi_{s}(0) } 
    + \cdots       \nonumber   \\
\psi_{\downarrow}(z_{1}) \psi_{\uparrow}( \bar{z}_{2} ) & = & - e^{-i \sqrt{2} \Phi_{c}(0) } 
    + \cdots           
\label{pairingnfl}
\end{eqnarray}

     The P-H symmetry interchanges the pairing and spin operators in
   the $ \uparrow \downarrow $ and $ \downarrow \uparrow $ channels. 

   From Eq.\ref{pairingnfl}, we can identify the pairing operators
\begin{equation}
 {\cal O}_{s}= e^{-i \sqrt{2} \Phi_{s}(0) },~~~ {\cal O}_{c}= e^{-i \sqrt{2} \Phi_{c}(0)},~~~
 {\cal O}_{\downarrow}= \partial \Phi_{\downarrow}(0) 
\end{equation}

  The paring operators in all the channels except in the $ \uparrow \uparrow $ channel
  have scaling dimension 1, therefore their correlation functions decay as
  $ \tau^{-2} $. Comparing these pairing operators with those at the FL fixed point
  ( Eq. \ref{pairingfl} ) to be discussed in the following, we find the pairng
  susceptibility in $ \downarrow \downarrow $ channel is enhanced. However,
 in contrast to the 2CK fixed point \cite{powerful}, the enhancement is so weak that
   there is {\em no} pairing susceptibility
  {\em divergence} at the impurity site in {\em any spin channel}.
 This result is somewhat surprising.  Naively, we expect pairing
 susceptibility divergence because the impurity interacts with the pairing density of the
 conduction electrons at the impurity site. However, the above explicit calculations showed
 that this is {\em not} true if there is only {\em one } channel of conduction electrons.
 Naively, we do {\em not} expect pairing susceptibility divergence in the 2CK, because
 the impurity spin interacts only with the total {\em spin } currents of channel 1 and 2, 
 {\em no } isospin currents of channel 1 and 2 are involved in the interaction. However,
 the explicit calculation of the 2CK showed that the pairing operator in the spin and flavor
 singlet channel has dimension 1/2 ( however, the  pairing operators in flavor singlet and spin triplet channel
 has dimension 3/2 ), therefore the spin and flavor singlet pairing susceptiblity at the impurity site
 is {\em divergent} \cite{powerful}. This indicates that we can achieve the pairing susceptiblity
 divergence without a pairing source term. We conclude that {\em more than } one channel of conduction
 electrons are needed to achieve the pairing susceptipility  {\em divergence}.

\subsection{ Fixed pointed 2}

   The boundary conditions are
\begin{equation}
  \psi_{\uparrow,L} = -\psi_{\uparrow,R},~~~
  \psi_{\downarrow,L} = -\psi^{\dagger}_{\downarrow,R}
\end{equation}

   The above boundary conditions can be expressed in terms of bosons
\begin{equation}
  \Phi_{\uparrow,L} =\Phi_{\uparrow,R} + \pi,~~~
  \Phi_{\downarrow,L} = -\Phi_{\downarrow,R} 
\end{equation}
 
 This fixed point is stable only when $\lambda=\alpha_{z}=0 $.
 If we define the P-H transformation $ \psi_{\uparrow} \rightarrow \psi_{\uparrow},
 \psi_{\downarrow} \rightarrow -\psi^{\dagger}_{\downarrow} $, then the spin and isospin currents
 transform as $ I^{x} \rightarrow -J^{x}, I^{y} \rightarrow -J^{y}, I^{z} \rightarrow J^{z};
  J^{x} \rightarrow -I^{x}, J^{y} \rightarrow -I^{y}, J^{z} \rightarrow I^{z}$.
 The Hamiltonian \ref{com} has this P-H symmetry if $\lambda=\alpha_{z}=0 $.

  This is the same fixed point as fixed point 1.

\subsection{ Fixed pointed 3}
   The boundary conditions are
\begin{equation}
  \psi_{\uparrow,L} =-\psi_{\uparrow,R},~~~
  \psi_{\downarrow,L} = -\psi_{\downarrow,R}
\end{equation}
   The above boundary conditions can be expressed in terms of bosons
\begin{equation}
  \Phi_{\uparrow,L} =\Phi_{\uparrow,R} + \pi,~~~
  \Phi_{\downarrow,L} = \Phi_{\downarrow,R}  + \pi
\end{equation}

   Both spin $\uparrow $ and $\downarrow $ electrons suffer $ \frac{\pi}{2} $
 phase shift. The physical picture is that the impurity spin is either totally
 screened by the spin current or the isospin current of conduction
 electrons depending on which coupling is stronger \cite{exchange}.
 This is a FL fixed point with $ O(4) $ symmetry. The finite size spectrum
  is listed in Table \ref{compactfl}. 

  The bosonized forms of the 4 leading irrelevant operators are \cite{another}
\begin{eqnarray}
  \widehat{a}(\tau) \frac{\partial \widehat{a}(\tau)}{\partial \tau}
 & = & \cos 2\Phi_{\downarrow}-\frac{1}{2} (\partial \Phi_{\downarrow}(0))^{2}  \nonumber   \\
  \widehat{b}(\tau) \frac{\partial \widehat{b}(\tau)}{\partial \tau}
 & = & -\cos 2\Phi_{\downarrow}-\frac{1}{2} (\partial \Phi_{\downarrow}(0))^{2}  \nonumber   \\
\widehat{a}\widehat{b}a_{\uparrow}(0)b_{\uparrow}(0) & = &   \partial\Phi_{\downarrow}(0,\tau)
  \partial\Phi_{\uparrow}(0,\tau)   \nonumber   \\
  a_{\uparrow}(0,\tau) \frac{ \partial a_{\uparrow}(0,\tau)}{\partial \tau} & + &
  b_{\uparrow}(0,\tau) \frac{ \partial b_{\uparrow}(0,\tau)}{\partial \tau}
  = (\partial \Phi_{\uparrow}(0,\tau))^{2}   
\label{last}
\end{eqnarray}

  The first order correction to the spin $\uparrow $ electron L-R Green function due to the 4th
  operator in Eq.\ref{last} is also given by Eq.\ref{many}. The correction due to the 3rd operator
  in Eq.\ref{last} can be similarly evaluated. We get the low temperature expansion
 of the spin $\uparrow $ electron conductivity
\begin{equation}
   \sigma_{\uparrow}(T) \sim \sigma_{u}(1+ T^{2} + T^{4} + \cdots)  
\label{phyup}
\end{equation}

   The first order correction to the spin $\downarrow $ electron L-R Green function  
  due to the first operator in Eq. \ref{last} is
\begin{eqnarray}
  \int d\tau \langle e^{- i \Phi_{\downarrow}( x_{1}, \tau_{1} )} \cos 2\Phi_{\downarrow}(0,\tau)
     e^{ i ( \Phi_{\downarrow}( x_{2}, \tau_{2} ) +\pi) }\rangle      \nonumber  \\
  -\frac{1}{2} \int d\tau \langle e^{- i \Phi_{\downarrow}( x_{1}, \tau_{1} )}
       ( \partial \Phi_{\downarrow}(0,\tau) )^{2}
     e^{ i ( \Phi_{\downarrow}( x_{2}, \tau_{2} ) +\pi)}\rangle    
\end{eqnarray}

  By using the following OPE:
\begin{eqnarray}
 : e^{-i \Phi_{\downarrow}( z_{1} )}: : e^{i  \Phi_{\downarrow}( z_{2})}: =
 (z_{1}-z_{2})^{-1}-i :\partial \Phi_{\downarrow}(z_{2}):  
                            \nonumber   \\
 +\frac{z_{1}-z_{2}}{2} :\partial^{2} \Phi_{\downarrow}(z_{2}):
 -\frac{z_{1}-z_{2}}{2} : (\partial \Phi_{\downarrow}(z_{2}) )^{2}: + \cdots
\end{eqnarray}

   It is ease to see that the  first integral vanishes and the  second are 
   the same as Eq.\ref{many}. The corrections due to the 2nd and the 3rd operators in Eq.\ref{last}
   can be similarly evaluated, the low temperature expansion
   of the spin $\downarrow $ electron conductivity follows
\begin{equation}
   \sigma_{\downarrow}(T) \sim \sigma_{u}(1+ T^{2} + T^{4} + \cdots)  
\label{phydown}
\end{equation}

   Note that only at the FL fixed point, the spin $ SU(2) $ symmetry is restored, 
  therefore the expansion coefficients in Eqs.\ref{phyup},\ref{phydown}
  are {\em different}.

     The total conductivity is the summation of the two spin components
\begin{equation}
  \sigma(T)= \sigma_{\uparrow}(T)+ \sigma_{\downarrow}(T)
   \sim 2 \sigma_{u}(1+ T^{2}+ T^{4} + \cdots   )
\end{equation}

  The boundary OPE of the spin and density of the {\em conduction electrons} are
\begin{eqnarray}
\psi^{\dagger}_{\uparrow}(z_{1}) \psi_{\uparrow}( \bar{z}_{2} ) & = & 
    ( z_{1}-\bar{z}_{2} )^{-1} +i \partial \Phi_{\uparrow} + \cdots     \nonumber   \\
\psi^{\dagger}_{\downarrow}(z_{1}) \psi_{\downarrow}( \bar{z}_{2} ) & = &
    ( z_{1}-\bar{z}_{2} )^{-1} +i \partial \Phi_{\downarrow} + \cdots     \nonumber   \\
\psi^{\dagger}_{\uparrow}(z_{1}) \psi_{\downarrow}( \bar{z}_{2} ) & = & e^{i \sqrt{2} \Phi_{s}(0) } 
    + \cdots       \nonumber   \\
\psi^{\dagger}_{\downarrow}(z_{1}) \psi_{\uparrow}( \bar{z}_{2} ) & = &  e^{-i \sqrt{2} \Phi_{s}(0) } 
    + \cdots         
\label{sdfl}
\end{eqnarray}

  The boundary OPE of the spin singlet and triplet pairing operators are
\begin{eqnarray}
\psi_{\uparrow}(z_{1}) \psi_{\uparrow}( \bar{z}_{2} ) & = & 0
    + \cdots       \nonumber   \\
\psi_{\downarrow}(z_{1}) \psi_{\downarrow}( \bar{z}_{2} ) & = & 0 +\cdots       \nonumber  \\
\psi_{\uparrow}(z_{1}) \psi_{\downarrow}( \bar{z}_{2} ) & = & -e^{-i \sqrt{2} \Phi_{c}(0) } 
    + \cdots       \nonumber   \\
\psi_{\downarrow}(z_{1}) \psi_{\uparrow}( \bar{z}_{2} ) & = &  e^{-i \sqrt{2} \Phi_{c}(0) } 
    + \cdots           
\label{pairingfl}
\end{eqnarray}

  The above equations should be compared with the corresponding Eqs.\ref{sdnfl} and \ref{pairingnfl}
  at the NFL fixed point.

\section{Conclusions}

  By the detailed discussions on the low temperature properties of the two related, but
 different single impurity models, we clarify the confusing conjectures and
 claims made on these two models.  In evaluating the single
 particle Green functions and pairing susceptibilities, all the degree of freedoms
 have to be taken into account, even though some of them decouple from the interactions
 with the impurity.
 We explicitly demonstrate that different quantum impurity models are simply 
 free chiral bosons with different boundary conditions. In Ref.\cite{sf}, the author 
 studied another single impurity model where the impurity couples to both
 the spin and the flavor currents of the two channel electrons ( 2CSFK). In Ref.\cite{hopping},
 the author solved a two level tunneling model which can also mapped to a single impurity model.
 As shown in Ref.\cite{twoimp},
 finite number of impurity models can always mapped to a single impurity model.
 From the results of this paper and Refs.\cite{sf,hopping},
 we conclude that in clean, finite number of impurity models (1) FL behaviors
 are extremely robust, any perturbation in the flavor sectors will destroy the NFL
 behaviors.(2) due to the phase space arguments given in this paper and
 in Refs.\cite{sf,hopping},  it is very
 unlikely to find the NFL linear $ T $ bahaviour of the electron conductivity
 which was observed  in the certain heavy fermion systems \cite{linear} and in the normal
 state of high- $T_{c} $ cuprate superconductors. There are three possible ways to explain
 this experimental observation (1) disorder effects \cite{bhatt} (1) Kondo lattice model \cite{lattice}
 (3) near to some quantum phase transitions \cite{millis}, for example,  near the phase transition between
 the metallic spin-glass and disordered metal \cite{phase,sachdev}.

\centerline{\bf ACKNOWLEDGMENTS}
We thank D. S. Fisher, B. Halperin, A. Millis, N. Read for helpful discussions.
This research was supported by NSF Grants Nos. DMR 9630064, DMR9416910 and Johns Hopkins University.

\appendix

\section{ CFT analysis of the stable fixed point of the 2CFAK} 

  We can also analysis the stable fixed point from CFT. Without
  losing generality, supposing at this
  fixed point, the impurity is totally absorbed by the channel 1
  conduction electrons $ \vec{{\cal J}}_{1}(x) = \vec{J}_{1}(x)
      + 2 \pi \delta(x) \vec{S} $. Slightly away the fixed point,
 the channel 2 conduction electrons also couple to the impurity.
 It has been shown by the author there is only {\em one} leading
 irrelevant operator $ Q^{0}_{0} =
   \vec{{\cal J}}_{1}(0) \cdot \vec{{\cal J}}_{1}(0) $ even
  in the $ O(2) $ one channel Kondo model \cite{line}. 
 The Hamiltonian $ H = H_{1} + H_{2} + H_{12} $ is
\begin{eqnarray}
    H_{1} & = & \frac{1}{3} \int dx 
  \vec{{\cal J}}_{1}(x) \cdot \vec{{\cal J}}_{1}(x) 
  + h \int dx {\cal J}^{3}_{1}(x) +
  \lambda_{1} \vec{{\cal J}}_{1}(0) \cdot \vec{{\cal J}}_{1}(0) + \cdots
                                         \nonumber     \\
    H_{2} & = & \frac{1}{3} \int dx 
  \vec{ J }_{2}(x) \cdot \vec{ J }_{2}(x) 
  + h \int dx J^{3}_{2}(x)           \nonumber   \\
   H_{12} & = & \lambda^{z}_{2} J^{z}_{2}(0) S^{z} + 
    \lambda_{2} ( J^{x}_{2}(0) S^{x} + J^{y}_{2}(0) S^{y} ) 
\end{eqnarray}

   At the fixed point $ \vec{S} \sim \vec{{\cal J}}_{1}(0)  + \cdots $, it
 is easy to show that the following 4 leading irrelevant operators with
   dimension 2 will be generated \cite{violate}
\begin{eqnarray}
  O_{1} & = & \frac{1}{3} ( \vec{ \cal{J}}_{1} (0) \cdot \vec{\cal{J}}_{1}(0)
  + \vec{J}_{2} (0) \cdot \vec{J}_{2}(0) )    \nonumber  \\
  O_{2} & = & \frac{1}{3} ( \vec{ \cal{J}}_{1} (0) \cdot \vec{\cal{J}}_{1}(0)
  - \vec{J}_{2} (0) \cdot \vec{J}_{2}(0) )    \nonumber  \\
  O_{3} & = & \frac{2}{3}  \vec{\cal{J}}_{1} (0) \cdot \vec{J}_{2}(0)
                            \nonumber   \\
  O_{4} & = & \frac{2}{3} {\cal J}^{3}_{1} (0) J^{3}_{2} (0) -\frac{1}{3} (
  {\cal J}^{1}_{1} (0) J^{1}_{2} (0)+ {\cal J}^{2}_{1} (0) J^{2}_{2} (0) )
\end{eqnarray}
  
  This four operators are consistent with those listed in Eq.\ref{four}.
 The first order perturbations in the leading irrelevant operators yield
 the typical fermi liquid behaviors:
 $ C_{imp} \sim \lambda_{1} \frac{\pi^{2}}{3} T ,
 \chi_{imp} \sim \frac{1}{2} ( \lambda_{1} + \frac{\lambda_{3}
 +\lambda_{4}}{3}) $. {\em In contrast to } one channel spin anisotropic
 Kondo model, $ R= 2(1 + \frac{\lambda_{3} + \lambda_{4}}{3 \lambda_{1} } ) $
 is non-universal.

  The second oder perturbations yield
\begin{equation}
  \sigma_{1}(T) \sim \sigma_{u} (1 + T ^{2} + \cdots ), ~~~
\end{equation}

    The residual conductivity of channel 2 comes from the potential scattering which
  is neglected in this appendix \cite{break}. The total conductivity is the summation of those
  from the two channels. Similar CFT analysis can be applied to 1CCK.

  The CFT analysis in this appendix can only show that the fixed point examined here
   is stable, but cannot
  rule out the possible existence of the other fixed points. The Abelian Bosonization
  analysis in Sec. II identified all the possible fixed points and showed this fixed 
  point is the only stable fixed point.

\begin{table} 
\begin{tabular}{ |c|c|c|c| } 
 $ O(4) $ & $ O(4)$ & $ \frac{l}{v_{F} \pi}( E-\frac{1}{4}) $  & Degeneracy  \\  \hline
    R     &    NS      &        0                               &    4      \\  
    NS    &    R       &        0                              &    4      \\    \hline
    R     &    NS+1st  &       $ \frac{1}{2} $                    &    16      \\ 
  NS+1st  &    R       &       $ \frac{1}{2} $                     &    16      \\ \hline
  R+1st   &    NS      &           1                            &    16       \\
    R     &    NS+2nd  &           1                            &    24      \\   
   NS     &    R+1st   &           1                            &    16      \\
   NS+2nd &     R      &           1                            &    24      \\  
\end{tabular}
\caption{ The finite size spectrum at the stable FL fixed point of the 2CFAK 
 with the symmetry $ O(4) \times O(4) $. See Ref.\cite{powerful} for explanations.
 It is the superposition of the finite size spectrum of free electrons
 and that of electrons  with phase shift $\pi/2 $.}
\label{flavor}
\end{table}

\begin{table} 
\begin{tabular}{ |c|c|c| } 
 $ O(4) $  & $ \frac{l}{v_{F} \pi}( E-\frac{1}{4}) $  & Degeneracy  \\  \hline
    R      &        0                               &    4      \\   \hline 
   R+1st   &        1                               &    16       \\  \hline
   R+2nd   &        2                               &    40      \\   
\end{tabular}
\caption{ The finite size spectrum at the FL fixed point of the 1CCK 
 with the symmetry $ O(4) $.} 
\label{compactfl}
\end{table}

\begin{table} 
\begin{tabular}{ |c|c|c|c| } 
 $ O(1) $ & $ O(3)$ & $ \frac{l}{v_{F} \pi}( E-\frac{3}{16}) $  & Degeneracy  \\  \hline
    R     &    NS      &        0                               &    2      \\  \hline
    NS    &    R       &       $ \frac{1}{8} $                    &    2      \\ \hline
    R     &    NS+1st  &       $ \frac{1}{2} $                    &    6      \\ \hline
  NS+1st  &    R       &       $ \frac{5}{8} $                     &    2      \\ \hline
  R+1st   &    NS      &           1                            &    2       \\
    R     &    NS+2nd  &           1                            &    6      \\   \hline
   NS     &    R+1st   &       $ 1+\frac{1}{8}$                   &    6      \\  \hline
   NS+2nd &     R      &       $ 1+\frac{1}{8}+\frac{1}{2} $                   &    2      \\  
\end{tabular}
\caption{ The finite size spectrum at the NFL fixed point of the 1CCK
 with the symmetry $ O(1) \times O(3) $.}
\label{compactnfl}
\end{table}


\begin{references}
\bibitem{semi} Strictly speaking, it should be a semi-product.
\bibitem{atten} The difference between $ \psi_{s,i} $ and $ \psi_{s} $  are neglected
                 in the following
\bibitem{ising} By $ O(1) $, we mean the Ising sector with $ c=1/2 $.
\bibitem{above} If keeping $ \gamma_{1}=\gamma_{2}=1 $, we get the weak 
                coupling RG flow Eq.\ref{danger}.
\bibitem{cut} The superscript $ H $ is omitted in the following without causing confusions.
\bibitem{another} The 3rd operator in the following equation will also generate
    dimension 2 operator $:(\partial \Phi_{sf}(0,\tau))^{2}: $, but this operator
    is just the linear combination of the first two.
\end{references}
\end{document}